\documentclass[12pt]{article} 
\usepackage{epsfig}
\begin{document}
\sloppy
\begin{titlepage}
\title{On the elastic behavior of a single polyelectrolyte chain}
\author{P. Haronska\footnote{e-mail: haronska@mpip-mainz.mpg.de},
  J. Wilder\footnote{e-mail: wilder@mpip-mainz.mpg.de}  
and
 T.A. Vilgis \footnote{e-mail: vilgis@mpip-mainz.mpg.de}
\\Max-Planck-Institut f\"ur Polymerforschung, \\Postfach 3148, D-55021 Mainz, Germany
}
\date{\today}

\end{titlepage}
\maketitle
\newpage
\begin{abstract}
This paper discusses the elastic behavior of a single polyelectrolyte chain.
A simple scaling analysis  as in self avoiding walk chains are not possible,
because three interplaying relevant length scales are involved, i.e., the 
Debye screening length and the Pincus blob size.  Therefore a selfconsistent
computation of an effective variational propagator is employed.
It is shown that the elastic force as a function $f$ of the 
distance $R$ behaves as $f \propto R$  for
small $f$. For larger forces we find a new regime,
characterized by deformations larger than a computed
electrostatic "blob size". These results are supported
by simulations and intuitive physical arguments.
\end{abstract}
short title: Elasticity of polyelectrolyte chains
\newpage

\section{Introduction}
In contrast to neutral
polymers polyelectrolytes are bearing ionizable groups, that
are able to dissociate into charged polymer chains and small counter-ions.
The interest in polyelectrolytes reaches back to the early days of polymer
science (see e.g. \cite{tanford}), due to their fundamental importance in
biology, biochemistry as well as in industrial applications. Proteins,
nucleic acids and superabsorber materials are only a few examples out of
the wide range of practical interest in polyelectrolytes. 
Nevertheless they belong 
to the least understood systems in macromolecular science \cite{schmitz}.
The reason why they are much less understood than for example neutral polymers
lies in the difficulty to apply renormalization group theories and scaling
ideas to systems in which long range (i.e. coulomb) forces are present, which
means that the range of the forces is of the order of the polymer size.

For many applications knowledge about the elastic behavior of
polyelectrolytes is of fundamental importance. One of them is, for example,
to understand the elastic
behavior of superabsorber materials, i.e.,  polyelectrolyte networks and
their thermodynamic properties. Most important amongst them is naturally the
swelling behavior. 

The elastic and the thermodynamic properties of classical networks
formed  by crosslinking of neutral polymer chains have been considered
by simple and successful theories \cite{Ltreloar}. The success of these
theories lies in the fact that most of them are 
based on the assumption that the elasticity of
the entire network is roughly given by the elasticity of a representative
single chain in the cross-linked network (see for example
\cite{Lreview}, \cite{Lrreview} for reviews). Many theories including
those which consider entanglements, work very well
for dry networks, i.e., networks that are not swollen by a good
solvent. The case where the networks are at
equilibrium swelling at the so-called $c^*$ point, when
the blob size is precisely identical to the
radius of gyration of the mesh \cite{Lggennes}  is
especially interesting. In this case the network elasticity can be
computed by the knowledge of the 
conformational behavior of a single swollen
chain \cite{Ralf}.

Naturally these simple theories are suitable for many purposes
but are, of course on the other hand, 
unable to answer difficult questions such as
the liquid solid transition, etc \cite{Gzipp, GCzipp}, but this 
is not the subject of the present paper. 
We would like to address another important issue of the subject
of network elasticity. In all theories the effects of 
interactions have been neglected. In the classical theories it was
assumed by purpose of mathematical simplicity that any interactions
between the chains to do not contribute to the elastic properties. This
is most likely valid for neutral polymer networks, because in dense
systems the excluded volume forces between the chains are screened to 
a large extend. In polyelectrolyte networks this is definitely not the 
case, because strong interactions between monomers rule the conformational
behavior \cite{Krems}. Indeed little is known about the 
interplay of elasticity (conformation) and strength of the interactions
in the theory of network elasticity. 

To get a first insight and physical feeling for the interplay of these effects, 
it is useful and legal to consider
single chain elasticity in weakly charged and 
flexible polyelectrolyte chains, although
the simple generalizations as carried out for neutral networks are
not promising. On the other hand simple scaling considerations
as done for excluded volume chains \cite{Lggennes,Lpincus} cannot
be employed {\em a priory}, because several independent length scales
are involved. The first is (apart from the chain length) the range
of the electrostatic interaction. For simplicity we use here 
a screened Debye potential $V({\bf r}) \propto 1/r \exp(-\kappa r)$, where
$\lambda_{\rm D} = 1/\kappa$ defines the 
screening length of the electrostatic
interactions. The other scale, in neutral SAW polymers introduced as
Pincus blob size $\xi_{\rm P}$ via the relation $f=k_{\rm B}T/\xi_{\rm P}$,
defines the elastic properties as a function of the excluded volume
interaction by the assumption that within scales $\xi_{\rm P}$
the chains shows excluded volume behavior. $f$ is the 
external force acting at the chain ends. (Since
we assume $f =$ constant, the same force acts on each
segment along the chain). Similar treatment is here
not possible, because  the interplay between
$\lambda_{\rm D}$ and $\xi_{\rm P}$ is unknown. It has been shown only
recently, how a field theory can be set up for the critical
behavior, i.e., chain length $N \to \infty$
of undeformed polyelectrolyte chains \cite{tanniestapp}.

As a first step we calculate the elastic response of a 
single polyelectrolyte
chain in solvent. Therefore we apply an 
constant external force on the ends of the chain, which corresponds
to the application of an external field. Here we follow the
simple idea in classical rubber theory, but determine the 
new propagator of the extended {\em interacting} chain.
To do so, we employ a variational principle and calculate the effective
propagator of the chain, which allows statements about the influence of 
conformation
{\bf and} interactions on the elasticity. 
Although this procedure seems to be oversimplified for
several reasons. The first is that it cannot be expected that polyelectrolyte
networks can be described by effective single chain models as in networks with
short range interactions. The second reason is that fluctuations are expected
to play an important role. Thirdly, the Debye-H\"uckel approximation for the
electrostatic potential does not seem to be sufficient. This can be seen in
\cite{borsali} where the Debye-H\"uckel potential simply appears by a random
phase approximation in a Gaussian density functional theory. Moreover there is
great evidence from numerical simulation that the Debye-H\"uckel approximations
fails under certain circumstances \cite{kremer}. 
However, the use of the Debye approximation together with the
assumption that the non interacting chain is Gaussian limits the 
model to weakly charged and flexible polyelectrolytes.
Nevertheless, despite these drawbacks, we expect first
principal result from the theory below. 
As such a result we derive the elastic
modulus of the single polyelectrolyte chain and compare it the the well known
cases, i.e., free chain and excluded volume chain. As a side remark 
we mention that the same method can be used to re-derive the Pincus
blob model for pure excluded volume chains. In this case the renormalized
propagator can be determined in a field theoretic $4-d$ - expansion
\cite{harrov}
 
The starting point of the calculation is the Green function of the free chain
without interactions. The Green function can be viewed as the Fourier -
Laplace transform of the distribution function for the chain ends. It will be
shown that the 
force in the system is treated through the analytic continuation of the 
Fourier transformed Greens function to the complex plane as will be shown in
section two. After having introduced a field theory the problem is mapped on
a Gaussian field theory with a propagator that formally in the Fourier space
can be written down exactly by making use of the proper self energy. According
to the well known Feynman variational inequality the 
sum of the Gaussian free energy and 
the mean-value of the interacting potential has to be minimized with respect 
to the proper self energy, which is our variational parameter. This leads to 
a non-linear integral equation for the proper self energy, which can be solved
systematic approximations.

The result of this consideration is that the chain is stretched in the long
ranged limit proportional to the applied force parallel to the force and is 
pushed proportional to the force perpendicular to it. 
This result is in good agreement with the simulation, that is also
presented in this paper. Moreover we support the computations by intuitive
physical arguments. The variational calculation and simulation indeed support
a scaling picture of the deformation of the polyelectrolyte chain.

The paper is organized as follows. In the next section we present the
mathematical model and provide the definitions. In section 3 we introduce a
field theory and calculate the variational equation for the proper self
energy.
In section 4 this variational equation is solved approximately. The results 
of the analytic calculation are presented in section 5. The results of the
computational simulation are presented in section 6. Section 7 is provided
with some scaling considerations.

\section{Model}

Let us first introduce the standard model which is employed here. Since we
restrict ourselves to flexible chains which are weakly
charged the Edwards model is the appropriate tool.
The starting point is thus the continuum version of
the dimensionless Edwards Hamiltonian  \cite{Ledwards}
\begin{eqnarray}
&\displaystyle\label{model1}   \beta H_{E} [{\bf r};{\bf f}] = {3 \over 2l^{2} } \int^{N_{0}}_{0} \mbox{d} s \
\left({\mbox{d} {\bf r} \over \mbox{d} s}\right)^2 + \beta
 \int^{N_{0}}_{0} \mbox{d} s \  {\bf f} {\mbox{d} {\bf r} \over \mbox{d} s}
& \nonumber \\&\displaystyle
 + \
\frac{bz^{2}}{2} \int^{N_{0}}_{0} \mbox{d} s \int^{N_{0}}_{0} \mbox{d} s' \
{\exp\left\{-\kappa\left \vert {\bf r} (s) - {\bf r} (s') \right \vert \right \}
\over \left \vert {\bf r} (s) - {\bf r} (s') \right \vert}\,,
\end{eqnarray}
where ${\bf r}(s)$ represents the chain conformation in three dimensions as a function of the
contour variable $s$, $b=e^{2}/4\pi
\epsilon_{0}\epsilon_{r}k_{B}T$ is the Bjerrum-length, $\beta$ is
$(k_{B}T)^{-1}$, where $k_{B}$ is the Boltzmann constant and $T$ denotes the absolute temperature. $l$ is the Kuhn segment length,
$z$ is the monomer charge in units of $e$, $\epsilon_{0}$ is the dielectric
constant and $\epsilon_{r}$ the relative dielectric constant. $N_{0}$ stands for the bare number of
monomers on the chain, ${\bf f}$ is the external force and $\kappa^{-1}$
denotes the Debye-H\"uckel screening length.

The correlation function can be calculated in terms of a path integral
\cite{Ledwards,Lfeynman,Lfreed} as follows:
\begin{eqnarray}
&\displaystyle\label{model2}
G\left({\bf r},N_{0};{\bf f}\right) =
\int^{{\bf r}(N_{0}) = {\bf r}}_{{\bf r}(0) = {\bf 0}}
{\cal D} {\bf r}(s) \
\exp\left\{- \beta H_{E}[{\bf r};{\bf f}]\right\}&.
\end{eqnarray}
Its Fourier transform is defined by
\begin{eqnarray}
&\displaystyle\label{model3}
G({\bf k},N_{0};{\bf f}) = \int^{ }_{ } \mbox{d}^3 {\bf r}
 \
\exp\{-i{\bf k} {\bf r} \} G\left({\bf r},N_{0};{\bf f}\right)&.
\end{eqnarray}
The averages of the
force-size relationship $\langle{\bf R}^2_\parallel\rangle$ and
$\langle{\bf R}^2_\perp\rangle$,
where ${\bf R}_\parallel$ denotes the parallel component
with respect to ${\bf f}$ and ${\bf R}_\perp$ is the
corresponding perpendicular part, are then readily
calculated by the general formulae
\begin{eqnarray}
\langle{\bf R}^2_\parallel\rangle[{\bf f}]
=
- \left. { \partial^2/\partial k^2_{3} \ G({\bf k},N_{0};{\bf f})
\over G({\bf k},N_{0};{\bf f})}\right\vert_{{\bf k}={\bf 0}}\label{model4}&
\end{eqnarray}
and
\begin{eqnarray}
&\displaystyle\label{model5}
\langle{\bf R}^2_\perp\rangle[{\bf f}]
=
- \left. {\sum^{2}_{i=1} \partial^2/\partial k^2_i \ G({\bf k},N_{0};{\bf f})
\over G({\bf k},N_{0};{\bf f})}\right\vert_{{\bf k}={\bf 0}}&.
\end{eqnarray}
By analytic continuation of the Fourier space to the complex plane, the
correlation function $G({\bf k},N_{0};{\bf f})$ can also be written as the zero-force
correlation function $G({\bf k}-i\beta {\bf f},N_{0};{\bf f}={\bf 0})$.
Substitution of Eq.  \ref{model2} into Eq. \ref{model3} yields:
\begin{eqnarray}
 G({\bf k},N_{0};{\bf f})=\int^{ }_{ } \mbox{d}^3 {\bf r}\ 
\exp\{-i{\bf k} {\bf r} \}
\int^{{\bf r}(N_{0}) = {\bf r}}_{{\bf r}(0) = {\bf 0}}
{\cal D} {\bf r}(s)\ \exp \left \{-\beta H_{E}[{\bf r};{\bf f}]\right \}
\label{model10}
\end{eqnarray}
 For constant force ${\bf f}$
Eq. (\ref{model10})  can be rewritten as:
\begin{eqnarray}
 G({\bf k},N_{0};{\bf f})&=&\int^{ }_{ } \mbox{d}^3 {\bf r}\ 
\exp\{-i({\bf k}-i\beta {\bf f}) {\bf r} \}
\int^{{\bf r}(N_{0}) = {\bf r}}_{{\bf r}(0) = {\bf 0}}
{\cal D} {\bf r}(s)\ \exp \left \{-\beta H_{E}[{\bf r};{\bf f}={\bf 0}]\right \}
\nonumber \\
&=&  G({\bf k}-i\beta {\bf f},N_{0};{\bf f}={\bf 0})\label{model8}
\end{eqnarray}
Consequently, to get results for  $\langle{\bf R}^2_\parallel\rangle$ and
$\langle{\bf R}^2_\perp\rangle$, we only have to calculate $G({\bf k},N_{0})$ and continue the
first argument of $G$ to the complex plane.

\section{Field-theoretical formulation}
The Laplace transform of $G({\bf k},N_{0})$ with respect to $N_{0}$ is defined by
\begin{eqnarray}
\tilde{G}({\bf k},\mu_{0})=\int^{\infty}_{0}{\mbox d}N_{0}\,\exp\{-\mu_{0} N_{0}\}G({\bf
  k},N_{0})\label{ft1} 
\end{eqnarray}
The function $\tilde{G}({\bf k},\mu_{0})$ can be calculated by the introduction of de Gennes'
zero-component field theory (see for example \cite{PanRab})
\begin{eqnarray}
\tilde{G}({\bf k},\mu_{0})=\lim_{n \to 0}\int{\cal D}\vec{\psi}\,\psi_{1}({\bf
  k})\psi_{1}(-{\bf k})\exp\{-\beta H[\vec{\psi}]\}\label{ft2}
\end{eqnarray}
Here the field theoretical Hamiltonian $H[\vec{\psi}]$ is given by
\begin{eqnarray}
\beta H[\vec{\psi}]&=&\frac{1}{2}\int_{{\bf k}}\vec{\psi}(-{\bf k})\left[ \mu_{0}
  +\frac{l^{2}}{6}k^{2}\right]\vec{\psi}({\bf k})\label{ft3} \\&+&\frac{(2\pi)^{3}}{8}\int_{{\bf k_{1}},{\bf
    k_{2}},{\bf k_{3}},{\bf k_{4}}}\vec{\psi}({\bf k_{1}})\vec{\psi}({\bf k_{2}})U({\bf
  k_{1}}+{\bf k_{2}})\delta({\bf k_{1}}+{\bf k_{2}}+{\bf k_{3}}+{\bf
  k_{4}})\vec{\psi}({\bf k_{3}})\vec{\psi}({\bf k_{4}})\nonumber
\end{eqnarray}
where $\int_{{\bf k}}$ is an abbreviation for $\int d^{3}{\bf k}/(2\pi)^{3}$ and
 $U$ denotes the Debye-H\"uckel potential in units of $\beta^{-1}$.
In the Fourier space $\tilde{G}({\bf k},\mu_{0})$ can be written exactly
as 
\begin{eqnarray}
\tilde{G}({\bf k},\mu_{0})=\left(\mu_{0} +\frac{l^{2}}{6}k^{2}+\Sigma({\bf k})\right)^{-1}\label{ft4}
\end{eqnarray}
where $\Sigma({\bf k})$ denotes the proper self energy. 
We now consider an approximate correlation function $\tilde{{\cal G}}({\bf
  k},\mu_{0})$ with an approximate proper self-energy ${\cal M}({\bf k})$. 
\begin{eqnarray}
\tilde{{\cal G}}({\bf k},\mu_{0})=\left(\mu_{0} +\frac{l^{2}}{6}k^{2}+{\cal M}({\bf k})\right)^{-1}\label{ft4a}
\end{eqnarray}
Defining the Hamiltonian ${\cal H}$ by
\begin{eqnarray}
\beta {\cal H}[\vec{\psi}]=\frac{1}{2}\int_{{\bf k}}\vec{\psi}(-{\bf k})\tilde{{\cal
    G}}^{-1}({\bf
  k},\mu_{0})\vec{\psi}({\bf k})\label{ft6}
\end{eqnarray}
$\tilde{{\cal G}}({\bf k},\mu_{0})$ can  be calculated in the following way
\begin{eqnarray}
\tilde{{\cal G}}({\bf k},\mu_{0})=\lim_{n \to 0}\int{\cal D}\vec{\psi}\,\psi_{1}({\bf
  k})\psi_{1}(-{\bf k})\exp\{-\beta {\cal H}[\vec{\psi}]\}\label{ft7}
\end{eqnarray}
In this notation the well-known Feynman inequality is given by:
\begin{eqnarray}
F\le {\cal F}+\langle H-{\cal H}\rangle_{{\cal H}}\label{ft8}
\end{eqnarray}
where
\begin{eqnarray}
\langle\dots\rangle_{{\cal H}}=\lim_{n \to 0}\frac{\int{\cal
    D}\vec{\psi}\,\dots\exp\{-\beta {\cal H}\}}{\int{\cal
    D}\vec{\psi}\,\exp\{-\beta {\cal H}\}}\label{ft9}
\end{eqnarray}
is the mean-value and ${\cal F}$ the free energy with respect to ${\cal H}$.
 The right hand side of the inequality
(\ref{ft8}) has to be minimized with respect to 
${\cal M}$. ${\cal F}$ and $\langle H-{\cal H}\rangle_{{\cal H}}$ can be written 
in terms of the correlation function $\tilde{{\cal G}}({\bf k},\mu_{0})$: 
\begin{eqnarray}
\beta {\cal F} = -\ln{\cal Z} = -\frac{n}{2}V\int_{{\bf k}}\ln \tilde{{\cal G}}({\bf k},\mu_{0})\label{ft10}
\end{eqnarray}
As can be shown easily the second term of the right hand side of inequality (\ref{ft8}) is
\begin{eqnarray}
\beta \langle H-{\cal H}\rangle_{{\cal H}}&=&-\frac{n}{2}V\int_{{\bf k}}{\cal M}({\bf
  k})\tilde{{\cal G}}({\bf
  k},\mu_{0})+\frac{\pi bz^{2}n^{2}}{2\kappa^{2}}V\left(\int_{{\bf k}}\tilde{{\cal G}}({\bf k},\mu_{0})\right)^{2}\nonumber \\ 
&+&\pi bz^{2}nV\int_{{\bf k_{1}},{\bf k_{2}}}\frac{\tilde{{\cal G}}({\bf
  k_{1}},\mu_{0})\tilde{{\cal G}}({\bf k_{2}},\mu_{0})}{\kappa^{2}+({\bf
    k_{1}}+{\bf k_{2}})^{2}}\label{ft12}
\end{eqnarray}
The general minimization condition reads
\begin{eqnarray}
\frac{\delta}{\delta {\cal M}({\bf
    q})}({\cal F}+\langle H-{\cal H}\rangle_{{\cal H}})=0\label{ft13}
\end{eqnarray}
where $\delta/\delta{\cal M}({\bf q})$ denotes the functional derivative with
respect to ${\cal M}({\bf q})$. After inserting Eqs. (\ref{ft10}) and
(\ref{ft12}) into Eq. (\ref{ft13}) one obtains
\begin{eqnarray}
{\cal M}({\bf q})&=&\frac{2\pi bz^{2} n}{\kappa^{2}}\int_{\bf k}\frac{1}{\mu_{0}
  +\frac{l^{2}}{6}k^{2}+{\cal M}({\bf k})}\nonumber \\
&+&4\pi bz^{2}\int_{{\bf k}}\frac{1}{\left(\kappa^{2}+({\bf q}+{\bf k})^{2}\right)\left( \mu_{0}
  +\frac{l^{2}}{6}k^{2}+{\cal M}({\bf k})\right)}\label{ft14}
\end{eqnarray}
This is a non-linear integral equation for ${\cal M}({\bf q})$, which in the
following has to
be solved approximately, since the exact solution is unknown. At this point it
should be stressed, that Eq. (\ref{ft14}) represents the well known Hartree
approximation.

Another important and useful point is, that the exact proper self-energy
$\Sigma({\bf k},\mu_{0})$ is less than or equal to the approximate proper self-energy
${\cal M}({\bf k},\mu_{0})$. This can be shown as follows: First of all we
introduce an infinitesimal auxiliary real field $h({\bf r})$ in the field
theoretical Hamiltonians $H$ (Eq. (\ref{ft3})) and ${\cal H}$ (Eq. (\ref{ft6})).
\begin{eqnarray}
\beta H[\vec{\psi}] \mapsto \beta H[\vec{\psi}]+\int_{\bf k} h({\bf
  k})\psi_{1}(-{\bf k})\label{ft15}
\end{eqnarray} 
An analogous extension has to be done for ${\cal H}$, where $h({\bf k})$ is the
Fourier transform of $h({\bf r})$. Using this Hamiltonian the 
exact free energy $F$ becomes a functional of the auxiliary field $h$. Thus:
\begin{eqnarray}
F[h]=-\ln\left(\int{\cal D}\vec{\psi}\,{\mbox e}^{-\beta H-\int_{{\bf k}}h({\bf
      k})\psi_{1}(-{\bf k})}\right)\label{ft16}
\end{eqnarray}
Evaluating Eq. (\ref{ft16}) for small $h$ yields
\begin{eqnarray}
F[h]=F[0]-\frac{1}{2}\int_{\bf k}\left|h({\bf k})\right|^{2}\tilde{G}({\bf k},\mu_{0})+{\cal O}(h^{4})\label{ft17}
\end{eqnarray}
The validity of Feynman inequality is unaffected 
by the introduction of the auxiliary
field $h$ and becomes after having neglected ${\cal O}(h^{4})$
\begin{eqnarray}
F[0]-\frac{1}{2}\int_{\bf k}\left|h({\bf k})\right|^{2}\tilde{G}({\bf k},\mu_{0})\le {\cal F}[0]+\langle H-{\cal H}\rangle_{\cal H}-\frac{1}{2}\int_{\bf k}\left|h({\bf k})\right|^{2}\tilde{{\cal
  G}}({\bf k},\mu_{0})\label{ft19}
\end{eqnarray}
Inequality (\ref{ft17}) can be rewritten as
\begin{eqnarray}
\frac{1}{2}\int_{\bf k}\left|h({\bf k})\right|^{2}(\tilde{{\cal G}}({\bf
  k},\mu_{0})-\tilde{G}({\bf k},\mu_{0}))\le{\cal F}[0]+\langle H-{\cal H}\rangle_{\cal H}-F[0]\label{ft18}
\end{eqnarray}
In the limit $n \to 0$ the right
hand side of Inequality (\ref{ft18}) vanishes. 
Due to
the fact that
inequality (\ref{ft18}) holds for any field $h$, we get
\begin{equation}
\tilde{G}({\bf k},\mu_{0})\ge\tilde{\cal G}({\bf k},\mu_{0})\label{ft20}
\end{equation}
which is equivalent to
\begin{equation}
{\cal M}({\bf k})\ge\Sigma({\bf k})\label{ft21}
\end{equation}   
Thus  ${\cal M}({\bf k})$ is proven to be 
an upper bound for $\Sigma({\bf k})$. 

\section{Approximate solution for the proper self-energy}

To do the explicit calculation 
let ${\cal M}_{r}({\bf q})$ be ${\cal M}({\bf q})-{\cal M}({\bf 0})$. 
Then ${\cal M}_{r}({\bf q})$ is given by
\begin{equation}
{\cal M}_{r}({\bf q})=\int_{\bf k}\left[\frac{1}{\kappa^{2}+({\bf q}+{\bf k})^{2}}-\frac{1}{\kappa^{2}+{\bf k}^{2}}\right]\frac{4\pi b z^{2}}{
  \mu+\frac{l^{2}}{6}k^{2}+{\cal M}_{r}({\bf k})}\label{s1}
\end{equation}
where $\mu=\mu_{0} +{\cal M}({\bf 0})$. In order to simplify the integral in
Eq. (\ref{s1}) we make the following approximation, which is valid for small
$\kappa$: 
\begin{equation}
{\cal M}_{r}({\bf q})=\int_{\bf \left|k\right|\geq \kappa}\left[\frac{1}{({\bf q}+{\bf k})^{2}}-\frac{1}{{\bf k}^{2}}\right]\frac{4\pi b z^{2}}{
  \mu+\frac{l^{2}}{6}k^{2}+{\cal M}_{r}({\bf k})}\label{s1a}
\end{equation}
 Equation (\ref{s1a}) could be solved by means of an iteration procedure  
following the scheme
\begin{eqnarray}
{\cal M}^{(1)}_{r}({\bf q})=\int_{\bf \left|k\right|\geq \kappa}\left[\frac{1}{({\bf q}+{\bf k})^{2}}-\frac{1}{{\bf k}^{2}}\right]\frac{4\pi b z^{2}}{
  \mu+\frac{l^{2}}{6}k^{2}}\label{s2}
\end{eqnarray}
\begin{eqnarray}
{\cal M}^{(p)}_{r}({\bf q})=\int_{\bf \left|k\right|\geq \kappa}\left[\frac{1}{({\bf q}+{\bf
      k})^{2}}-\frac{1}{{\bf k}^{2}}\right]\frac{4\pi b z^{2}}{
  \mu+\frac{l^{2}}{6}k^{2}+{\cal M}^{(p-1)}_{r}({\bf k})}\label{s3}
\end{eqnarray}
but we show below that in the variational technique this procedure
is not necessary. {\em En effet}, it is show right below, that the 
one loop renormalization agrees with the first order perturbation in the
limits we investigate.

It can be seen from Eqs. (\ref{s2}) and (\ref{s3}) that ${\cal M}_{r}({\bf q})$
is actually a function of the dimensionless parameters ${\bf q}l$ and
$\kappa l$.
Denoting ${\bf Q}={\bf q}l/\kappa l$ and ${\bf K}={\bf k}l/\kappa l$ Eq.(\ref{s1})
becomes 
\begin{equation} 
{\cal M}_{r}({\bf Q}\kappa l)=\frac{1}{\kappa^{2} l^{2}}\int_{{\bf
  \left|K\right|}\geq 1}\left[\frac{1}{({\bf Q}+{\bf K})^{2}}-\frac{1}{{\bf
      K}^{2}}\right]\frac{4\pi \kappa b z^{2}}{
  \frac{\mu}{\kappa^{2} l^{2}}+\frac{1}{6}K^{2}+\frac{{\cal M}_{r}({\bf
      K}\kappa l )}{\kappa^{2} l^{2}}}\label{s4}
\end{equation}
In the limit of small $\kappa l$ the validity of the ansatz 
\begin{equation}
{\cal M}_{r}({\bf K} \kappa l)=\alpha K^{2} \kappa^{2}
l^{2}+{\cal O}((K\kappa l)^{4})\label{s4a} 
\end{equation}
can be checked using
Eqs. (\ref{s2}) and (\ref{s3}). Consequently, the only remaining
task is to calculate
$\alpha$ from Eq. (\ref{s4}) selfconsistently. Therefore we introduce in the
integral of Eq. (\ref{s4}) spherical coordinates $(K,\vartheta ,\varphi )$ and perform
the integration over $\varphi$ and $\vartheta$. The second derivative with
respect to $Q$ at $Q=0$ yields the following equation for the coefficient
$\alpha$:
\begin{eqnarray}  
\alpha =\frac{2\kappa b z^{2}}{3\kappa^{4} l^{4} \pi}\int_{1
  }^{\infty}{\mbox d}K\,\frac{1}{K^{2}}\frac{1}{
  \frac{\mu}{\kappa^{2} l^{2}}+\frac{1}{6}K^{2}+{{\cal M}_{r}(K \kappa l) 
\over \kappa^{2} l^{2}}} \label{s5}
\end{eqnarray}
We perform the integral by  only taking into account the most singular term
with
respect to  $\kappa$. The result of this calculation is according to Eq. (\ref{s4a})
\begin{equation}
\alpha=\frac{2bz^{2}}{3l^{2}\pi \mu \kappa}
\left [1+{\cal O}\left (\kappa l \over \sqrt{\mu}  \right)\right]\label{s6}
\end{equation}
It is important to note that Eq. (\ref{s6}) coincides exactly with the first-order
 term in the perturbation expansion. Therefore the higher order 
terms within the Hartree-approximation do not contribute to the coefficient $\alpha$. 
This underlines the quality of the first-order approximation.

The constant term of the approximate proper self-energy ${\cal M}({\bf 0})$ is
given by the expression
\begin{eqnarray}
{\cal M}({\bf 0})=
{2 bz^{2} \over \pi}\int^{\infty}_{ \kappa}{\mbox{d} k \over  \mu_{0}+
{\cal M}({\bf 0})
  +\frac{l^{2}}{6}k^{2}+{\cal M}_{r}(k)}\label{ftneu}
\end{eqnarray}
Since $\mu_{0}$ is a finite number greater than zero, the main contribution to
the integral in Eq. (\ref{ftneu}) comes from large $k$. Eq. (\ref{s1a})
shows, that $\lim_{q \to \infty} {\cal M}_{r}({\bf q})=-{\cal M}(\bf 0)$,
because of the fact, that the first term in the brackets of the integrand can be neglected in the considered limit. Then the right hand
side of Eq. (\ref{s1a}) is exactly minus the right hand side of
Eq. (\ref{ftneu}). Thus for large $k$ the integrand does not depend on ${\cal
  M}$ and therefore on $\kappa$. Consequently, $\mu=\mu_{0}+{\cal
  M}({\bf 0})$ contains no singularity for small $\kappa$.   



Note that  we have confined ourselves 
to small values of $q$ so far, i.e., 
such values for which $q \ll \kappa$ is satisfied.
This makes sense, if we only consider the end-to-end distance of the chain
without external force. In this paper, however, we introduce an external force
on the chain. Therefore the restriction  $q \ll \kappa$ may be to strong
according to Eqs. (\ref{model4}) und (\ref{model5}).

Starting from Eq. (\ref{s2}) we obtain
\begin{equation}
{\cal M}^{(1)}_{r}({\bf q})=\int_{\vert {\bf k}\vert \ge \kappa}
\left[\frac{1}{({\bf q}+
{\bf k})^{2}}-\frac{1}{{\bf k}^{2}}\right]\frac{4\pi b z^{2}}{
  \mu+\frac{l^{2}}{6}k^{2}}\label{app1}
\end{equation}
The integral in Eq. (\ref{app1}) can be calculated in a double expansion. The result is
\begin{equation} \label{app2}
{\cal M}^{(1)}_{r}({\bf q})
=
{b z^{2} \over  \pi} {\kappa \over \mu}
\left [1 +
{\cal O}\left (  {\kappa l\over \sqrt{\mu}}  \right)
\right]
\sum^{\infty}_{n=0} \
{1 \over (2n+1)(2n+3)} \left({q \over \kappa}\right)^{2n+2}
\end{equation}
If one neglects ${\cal O}\left ( \left ( \kappa^{2} l^{2} / \mu \right)^{3/2} \right)$
Eq. (\ref{app2}) becomes
\begin{equation} \label{app3}
{\cal M}^{(1)}_{r}({\bf q})
=
{b z^{2} \over 4 \pi} {\kappa \over \mu}
\left ( {q \over \kappa} - {\kappa \over q} \right )
\ln \left ({1 + q/\kappa \over 1 - q/\kappa} \right)
+ 
{b z^{2} \over 2 \pi} {\kappa \over \mu}
\end{equation}
As can be seen from Fig. (1)  the quadratic approximation  of 
${\cal M}^{(1)}_{r}$ works well even for values such that $q \approx \kappa$. It is very
important to notice that  ${\cal M}^{(1)}_{r}$ has an imaginary part as soon as
$q > \kappa$, which gives also the limit of the force $f$, since the the
vector ${\bf q}$ contains the external force as third component. Indeed we
need here the condition $\beta f/\kappa < 1$. 

An inspection of Eq. (\ref{app1}) shows that the appearance of 
the imaginary part is
clearly an artifact. 
Therefore terms of ${\cal O}(\kappa l/\sqrt{\mu})$ cannot be neglected 
even if $\kappa l/\sqrt{\mu}$  becomes very small. 
 For $q > \kappa$ the quantity
$\kappa l/\sqrt{\mu}$ cannot be used as a small
parameter within the perturbation expansion. 
As a consequence we expect a new regime which will be detected also in the
simulation and the scaling theory below.

\begin{figure}
\epsfig{file=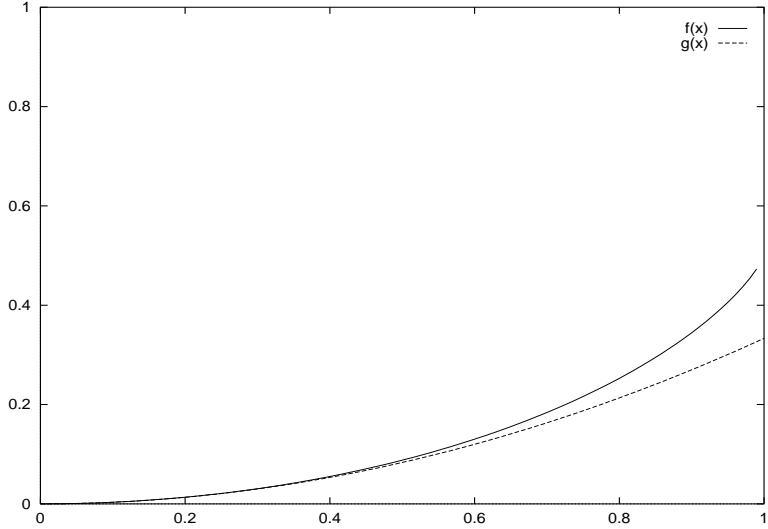,width=11cm,height=7cm}
\caption{Graphical illustration of the validity or our quadratic approximation
  for the proper self energy: quadratic approximation (dashed function), exact
  function according to Eq.(38) (bold-faced line)}
\end{figure}\label{bild1}

%
%

\section{Results}
Inserting the approximate result for the proper self-energy $\Sigma_{}({\bf k})$ into
Eq. (\ref{ft4}) yields an explicit expression for the correlation function
$\tilde{G}({\bf k},\mu)$ with the shifted chemical potential $\mu$ as
mentioned above
\begin{equation}
\tilde{G}({\bf k},\mu)=\left( \mu
  +\frac{l^{2}}{6}k^{2}+\frac{2bz^{2}}{3\pi  \mu \kappa}k^{2}\right)^{-1}\label{r1}
\end{equation}
Now the conformational free energy of the chain under the influence of a force
${\bf f}$ can be calculated very easily by
$\phi(\mu,{\bf f})=-\ln {\tilde G}({\bf 0},\mu;{\bf f})=-\ln {\tilde G}(-i\beta{\bf f},\mu)$. Using the
well-known thermodynamic relationship
\begin{eqnarray}
N=\frac{\partial \phi(\mu,{\bf f})}{\partial \mu}=-\frac{\partial \ln {\tilde G}(-i\beta{\bf f},\mu)}{\partial \mu}\label{r2}
\end{eqnarray}
we express $\mu$ depending on its conjugate variable $N$ and the force ${\bf
  f}$. Note that the variable $N$ is not the bare number of monomers $N_{0}$ since
$\mu$ is a shifted chemical potential, but $N$ is proportional to
$N_{0}$, indeed it is easily seen
that $N < N_0$. Considering only singular terms in $\kappa$ and neglecting terms of 
order $f^{4}$ this calculation yields:
\begin{eqnarray}
\mu=\frac{1}{N}+\frac{4Nbz^{2}\beta^{2}f^{2}}{3\pi \kappa}+{\cal O}(f^{4})\label{r2a}
\end{eqnarray}
Substituting Eq. (\ref{r2a}) into Eq. (\ref{r1}) we get $G({\bf k},N)$.
According to Eqs. (\ref{model4}) and (\ref{model5}) $\langle{\bf R}^2_\parallel\rangle$ and
$\langle{\bf R}^2_\perp\rangle$ can be calculated from Eq. (\ref{r2}). Expanding in a
power series for small forces to second order and again considering only
most singular terms  for small $\kappa$, 
$\langle{\bf R}^2_\parallel\rangle$ becomes for $\beta f/\kappa<1$
\begin{eqnarray}
\langle{\bf R}^2_\parallel\rangle[f]=\frac{4N^{2}bz^{2}}{3\pi
  \kappa}+\beta^{2} f^{2}\frac{8N^{4}b^{2}z^{4}}{9 \pi^{2} \kappa^{2}}+{\cal O}(\frac{\beta^{4}f^{4}}{\kappa^{4}})\label{r4}
\end{eqnarray}
 The square root of the mean square elongation  can
be written according to Eq. (\ref{r4}) as
\begin{equation}
\sqrt{\langle{\bf R}^2_\parallel\rangle[f]-\langle{\bf R}^2_\parallel\rangle[0]}\sim f\label{r6}
\end{equation}
which is a Hook-like law.

Making the same approximation as mentioned above the root mean square end-to-end
distance 
perpendicular  to the force $f$, $\sqrt{\langle{\bf R}^2_\perp\rangle}$, decreases with
$f$ for $\beta f/\kappa<1$, which is contrary to a Gaussian chain \cite{Lggennes}.
In
particular: 
\begin{equation}
\langle{\bf R}^2_\perp\rangle[f]=\frac{8N^{2}bz^{2}}{3\pi
  \kappa}-\beta^{2} f^{2}\frac{16N^{4}b^{2}z^{4}}{3 \pi^{2} \kappa^{2}}+{\cal O}(f^{4})\label{r5}
\label{r7}
\end{equation}
For  $f=0$
the 
perpendicular end-to-end distance becomes exactly twice  $\langle{\bf R}^2_\parallel\rangle[0]$.

\section{Simulation}

In this section we briefly
present the results from our Computer-simulations on a
single polyelectrolyte chain. Again the aim is to get a force-size
relationship for different values of the Debye-H\"uckel screening length
$\lambda_{D}$. Therefore the monomers of the chain are located on the lattice 
points of a simple cubic lattice, i.e., the Kuhn segment length is equivalent
 to the distance between two neighboring lattice sites. Because of the fact
 that we only consider
static properties of the chain, the algorithm of choice is clearly the
pivot-algorithm, where one randomly chooses a link in the chain and then
rotates this link together with the rest of the chain to a randomly chosen new
orientation of the lattice \cite{XYZbinder}. 
Whether this configuration will be excepted or not is decided by a simple
Metropolis-algorithm  \cite{XYZbinder}.

We consider a chain of $N=200$ monomers with Debye-H\"uckel interaction
between them. We made three different runs for the Debye-H\"uckel screening
length $\lambda_{D}=$ 5, 10 and 15 in units of the lattice constant.
\newline
As the initial configuration we have chosen a totally stretched chain on the 
lattice. After $10^{6}$ pivot-steps, we defined the starting 
configuration for the further simulation. In the case without an applied force
we made $1.6*10^{7}$ pivot steps to get the end-to-end distance of the 
forceless 
reference chain. On the other hand in the case with an applied force we made 
$4*10^{6}$ pivot steps. To get sufficient statistical independent values for
end-to-end distances we stored the configuration after every 8000 steps.
The results of the force-size relations are plotted in figure 2 and 3.
 
\begin{figure}
\epsfig{file=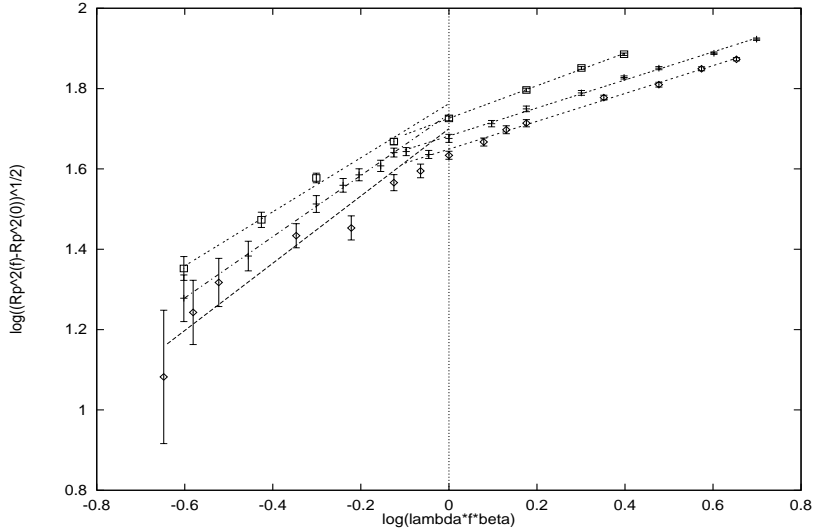,width=11cm,height=7cm}
\caption{Plot of the simulation data: Square root of the end-to-end distance
  parallel to the force minus the corresponding forceless case in double
  logarithmic scale}
\end{figure}

\begin{figure}
\epsfig{file=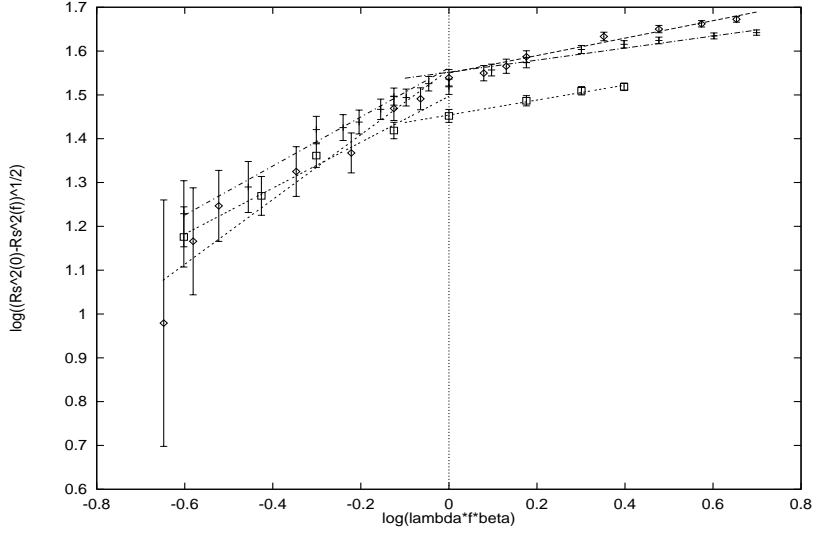,width=11cm,height=7cm}
\caption{Plot of the simulation data: Square root of the negative end-to-end distance
  perpendicular to the force minus the corresponding forceless case in double
  logarithmic scale}
\end{figure}

On this double logarithmic scale we find for each screening length
$\lambda_{D}$ two linear regimes. The change of the slope is clearly given in
the region $\lambda_{D}\beta f \approx 1$, which is in excellent agreement
with our theory. The slopes of the two regimes  and therefore
the exponent of the force $f$ in the force-size relationship are given in
table 1 for $\sqrt{\langle{\bf R}^{2}_{\parallel}\rangle}$ and in table 2 for
$\sqrt{\langle{\bf R}^{2}_{\perp}\rangle}$. These values are calculated by a linear
regression of the simulation data in figure 2 and 3.

\begin{table}
\begin{center}
\caption{Exponents for the force f in the force-size relationship for
  stretching of the chain parallel to the applied force}
\begin{tabular}{c|c|c|c}
 Exponent & $\lambda_{D}=5$&$\lambda_{D}=10$&$\lambda_{D}=15$\\
\hline
1. Regime&0.67&0.76&0.84 \\
\hline
2. Regime&0.4&0.35&0.34 \\
\hline
\end{tabular}
\end{center}
\end{table}

\begin{table}
\begin{center}
\caption{Exponents for the force f in the force-size relationship for
  stretching of the chain perpendicular to the applied force}
\begin{tabular}{c|c|c|c}
 Exponent & $\lambda_{D}=5$&$\lambda_{D}=10$&$\lambda_{D}=15$\\
\hline
1. Regime&0.52&0.56&0.74 \\
\hline
2. Regime&0.17&0.14&0.2 \\
\hline
\end{tabular}
\end{center}
\end{table}

As one can see from table 1 and 2 the exponents of the first regime increase
with increasing $\lambda_{D}$ for both cases stretching parallel and
perpendicular to the applied force ${\bf f}$. According to our theoretical
results we expect that the exponents for the first regime tend to one for
larger $\lambda_{D}$ which is in good agreement with our data in table 1 and
2. For the second regime we find a drastically lower exponents in both tables.
At this point it should be stressed that it is very difficult to get data for 
 $\lambda_{D}$ greater than about 15, because in this case the applied force
 ${\bf f}$ has to be so small that the transition from the first to the second
 regime lies within the numerical mistakes.

\section{Scaling considerations}

The analytical and numerical results suggest the following 
physical picture. At zero force, the polyelectrolyte 
chain is given by a chain
of blobs. The blob size is entirely determined by the 
electrostatic properties (see Eq. (\ref{r4})) .
The low force regime (see fig. 4) can be viewed by the 
picture that the already greatly elongated chain of blobs
becomes stretched. The chain size for zero force
is determined by 
\begin{equation}
R_0 \equiv \sqrt{\langle {\bf R}^2_{\parallel}\rangle} \simeq 
\left( \frac{bz^2}{\kappa}\right)^{1/2} N
\end{equation}
where we have ignored numerical prefactors. The fraction determined 
an effective step, or blob size, that is given by the 
charge $z$ and the Debye screening length $\kappa^{-1}$, i.e.,
$\xi_e \propto z \sqrt{b\lambda_{\rm D}}$. For the latter equation we have
assumed that the chain is weakly charged. Thus $\xi_{e}$ is consistent with
the assumptions and the use of the random walk chain model. Remember
that the variable $N$ is not the true degree of polymerization, but corresponds
to the  renormalized chain length via the relation
(steepest descent Laplace inversion) $N \simeq 1/(\mu_{0} + {\cal M}(0))$. 
This chain of blobs can become elongated until the chain of blobs becomes
fully stretched. 
In this low force regime the parts of the
chain inside the blobs do not take part on the deformation process.
(see fig. 4). This deformation process corresponds to the first
regime in the simulation, i.e., for forces $f < {\rm k_B}T/\lambda_{\rm D}=f_{c}$.

For larger forces ($f > f_{c}$) the blob size $\xi_e$ is no longer important, because 
the parts of the chain inside the blobs become deformed. This 
corresponds to a new Pincus regime, where the relevant blob
size is now given by $\xi_{\rm P} = {\rm k_B}T/f$. The simulation
clearly divides both regimes at forces $\lambda_{\rm D} = {\rm k_B}T/f$.
Note that this fact has been already
used in the analytical calculation above. The latter situation
is very similar to the case considered by Rabin and Alexander, when
the stretching of polymer brushes has been discussed \cite{rabin}. 
Thus the blob size becomes diminished
according to the idea pointed originally out by Pincus.
This can be seen clearly from the simulations. The Pincus
regime starts then at $R \sim \lambda_{\rm D}$.
\begin{figure}
\epsfig{file=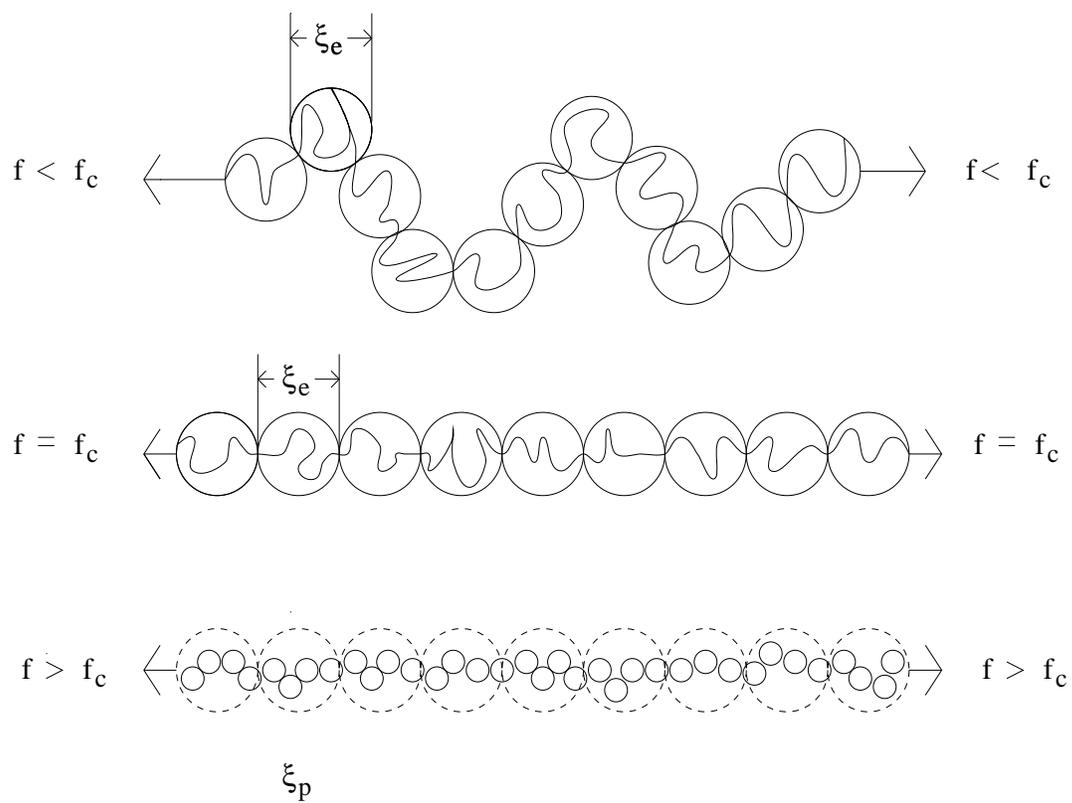,width=14cm,height=10.5cm}
\caption{Blob-model of a polyelectrolyte chain with different external forces
acting on the monomers (for details see text).}
\end{figure}

\section{Conclusions}

In the previous sections we analyzed the force-size relationship of a single
polyelectrolyte chain in a solvent. First we made theoretical considerations
on this problem. Then we compared our results with computer-simulation data and
some scaling considerations. In every three cases we find a transition from
one to another regime at forces which are about $\lambda_{D} \beta f \approx
1$ where $\lambda_{D}=1/\kappa$ the Debye-H\"uckel screening length in the
assumed Debye-H\"uckel potential. The most important result of the present
paper is the coupling between conformational degrees of freedom and the
interactions in the elastic response of a single chain. The force is still
proportional to the thermal energy $k_{B}T$, typical entropy elastic chains,
but the single chain modulus becomes strongly influenced by the interactions,
via the Debye screening length $\lambda_{\rm D} = \kappa^{-1}$.

Moreover the simulation data are in agreement with our theoretical
predictions,
that in the first regime the end-to-end distance of the chain depends
linearly on the applied force in the long range limit of the Debye-H\"uckel
potential, which means in solvents with low salt concentration. In a
subsequent paper we will extend this model and theoretical approach 
to the case of many crosslinked
chains \cite{wilder} and make further predictions on the elasticity of
polyelectrolyte networks.

\section*{Acknowledgments}

The authors wish to thank S. Stepanow, U. Micka, K. Kremer and M. P\"utz for
helpful discussions  and
 Firma Stockhausen Gmbh, D-47705 Krefeld, Germany for financial support.




\end{document}